\documentclass[aps,prl,letterpaper,superscriptaddress,twocolumn,showpacs,floatfix,10pt]{revtex4-1}
\usepackage{color}
\usepackage{ulem}
\usepackage{hyperref}
\usepackage{enumerate}
\usepackage{amsmath}
\usepackage{graphicx}

\def \be {\begin{equation}}
\def \ee {\end{equation}}
\def \ba {\begin{array}}
\def \ea {\end{array}}
\def \bea {\begin{eqnarray}}
\def \eea {\end{eqnarray}}

\def \ble {\begin{widetext}\begin{equation}}
\def \ele {\end{equation}\end{widetext}}
\def \blea {\begin{widetext}\begin{eqnarray}}
\def \elea {\end{eqnarray}\end{widetext}}

\def \nn {\nonumber}

\def \blea {\begin{widetext}\begin{eqnarray}}
\def \elea {\end{eqnarray}\end{widetext}}

\def \be {\begin{equation}}
\def \ee {\end{equation}}
\def \ba {\begin{array}}
\def \ea {\end{array}}
\def \bea {\begin{eqnarray}}
\def \eea {\end{eqnarray}}
\def \nn {\nonumber}

\def \p {\partial}

\def \and {{~\textrm{and}~}}

\begin{document}

\title{Duality of Ryu-Takayanagi surfaces inside and outside the horizon}

\author{Wu-zhong Guo}
\email{wuzhong@hust.edu.cn}
\affiliation{School of Physics, Huazhong University of Science and Technology,
Luoyu Road 1037, Wuhan, Hubei 430074, China}

\author{Jin Xu}
\email{xujin1@hust.edn.cn}
\affiliation{School of Physics, Huazhong University of Science and Technology,
Luoyu Road 1037, Wuhan, Hubei 430074, China}

\begin{abstract}
We study the Ryu-Takayanagi (RT) surfaces associated with timelike subregions in static spacetimes with a horizon. It is possible to find the analytical continuation of the RT surfaces that can extend into the horizon, allowing us to probe the interior of the black hole. The horizon typically divides the RT surface into two distinct parts. We demonstrate that the area of the surface inside the horizon can be reconstructed from the contributions of the surfaces outside the horizon, along with additional RT surfaces for spacelike subregions that are causally related to the timelike subregions. This result provides a concrete realization of black hole complementarity at the level of classical metric, where the spacetime in the black hole interior can be reconstructed from the degrees of freedom outside the horizon.

\end{abstract}

\maketitle

\section{Introduction}

The event horizon and singularity of a black hole are crucial probes of the quantum nature of gravity. Although information within the event horizon cannot be directly transmitted to the exterior, black hole complementarity suggests that the degrees of freedom in the interior are intimately linked to those in the exterior \cite{Stephens:1993an,Susskind:1993if}.

In the context of AdS/CFT, information within the black hole interior can be probed by observables in the dual field theories. Specifically, the bulk codimension-2 Ryu-Takayanagi (RT) surface, serves as a useful holographic probe, that is associated with the entanglement entropy (EE) of spacelike subregions on the boundary \cite{Ryu:2006bv}\cite{Hubeny:2007xt}. However, for RT surfaces anchored to spacelike subregions, the black hole horizon acts as a barrier \cite{Hubeny:2012ry}\cite{Engelhardt:2013tra}, limiting the extent of the RT surface. Nevertheless, RT surfaces corresponding to subregions on the two boundaries of an eternal black hole can probe the interior of a black hole \cite{Hartman:2013qma}. Other interesting holographic probes include correlation functions of heavy operators \cite{Fidkowski:2003nf,Frenkel:2020ysx} and complexity \cite{Stanford:2014jda,Brown:2015bva}.

Recently, it has been proposed that the concept of entanglement entropy can be extended to timelike subregions \cite{Doi:2022iyj}, which may be interpreted as pseudoentropy \cite{Nakata:2020luh}. In recent years there are many works on entanglement in time from various aspects \cite{Olson:2010jy}-\cite{Milekhin:2025ycm}. Timelike entanglement entropy can be well-defined and computed in quantum field theories (QFTs) through an analytical continuation of correlation functions. Although the physical meaning of entanglement in time is not yet fully understood, it is nevertheless believed that the RT formula could be applicable to timelike entanglement entropy in the context of AdS/CFT. However, a complete holographic proposal has yet to be established. Different proposals for holographic RT surfaces exist. In \cite{Doi:2022iyj, Doi:2023zaf}, the authors suggest that the RT surfaces for timelike EE are given by both spacelike RT surfaces and their timelike counterparts, while in \cite{Heller:2024whi}, it is argued that the RT surfaces correspond to extreme surfaces in complexified geometry. In the case of AdS$_3$, the two proposals yield equivalent results of timelike EE, but the distinction may emerge in higher-dimensional examples. Despite the incomplete understanding of holographic timelike EE, one intriguing feature is that the RT surfaces could extend into the black hole horizon, thus probing the interior, including the singularity. This could provide a novel probe to investigate the interior of a black hole \cite{Doi:2023zaf}\cite{Anegawa:2024kdj}. In \cite{Guo:2024lrr}, the authors find that timelike EE is related to spacelike EE in some examples of 2-dimensional CFTs, the holographic interpretation of this relation remains unclear.

\begin{figure}[htbp]
  \centering
  \includegraphics[width=0.5\textwidth]{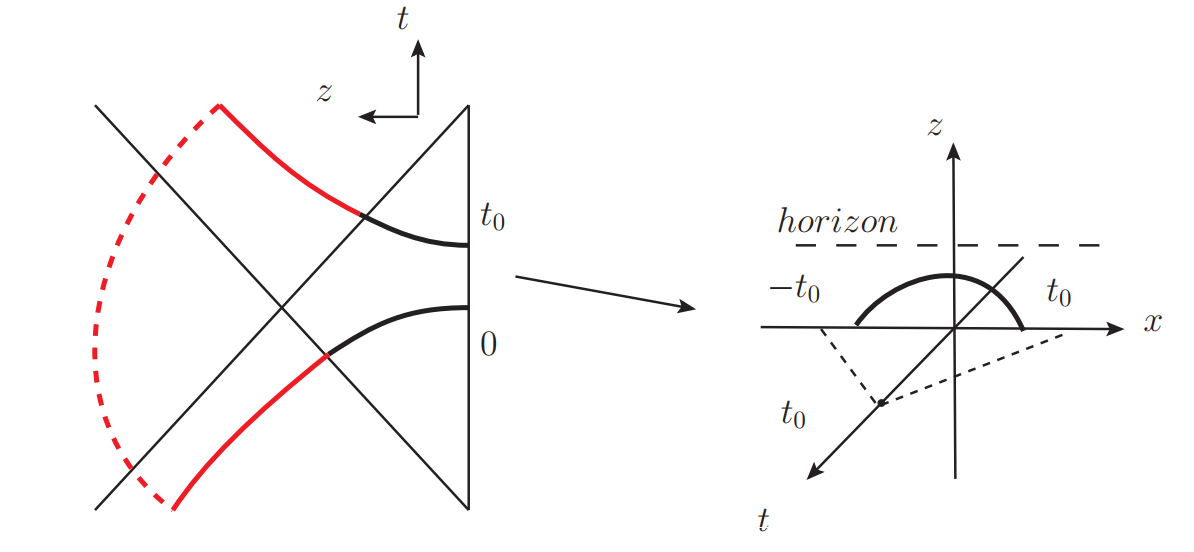}
  \caption{The black hole horizon divides the RT surface for timelike subregions into two parts. The coordinates $z$, $x$ and $t$ represent the holographic, spatial, and temporal directions, respectively. The red solid and dashed lines denote the portion of the RT surface inside the horizon, while the black solid line connecting the horizon to the boundary represents the RT surface outside the horizon. The duality relation states that the area of the RT surface inside the horizon equals the area outside the horizon, plus the contribution from RT surfaces for spacelike subregions on the Cauchy surface $t=0$ (as shown in the right panel).  }
 \label{general}
\end{figure}
The work presented in this Letter is motivated by recent studies of timelike EE and its holographic dual. We propose a method to construct RT surfaces for timelike subregions by analytically continuing their Euclidean counterparts, which correctly reproduces the expected results for timelike EE. This construction naturally leads to RT surfaces extending into a complexified geometry, consistent with the framework outlined in \cite{Heller:2024whi}. Importantly, the method is new and broadly applicable to general static backgrounds in arbitrary dimensions. Furthermore, we establish a relation between holographic timelike and spacelike EE for strip subregions in higher-dimensional backgrounds. This finding indicates that such a relation should hold in more general settings, extending beyond the specific two-dimensional CFTs case of \cite{Guo:2024lrr}.

More precisely, we will consider a static spacetime with a horizon. On the boundary, we choose a timelike strip subregion between the time interval $[0,t_0]$ (see Fig.\ref{strip} for an illustration). We would like to consider the extreme surfaces anchored to the boundary of the timelike subregion. These extreme surfaces for timelike subregions generally extend into the horizon and approach the singularity of the black hole. However, the RT surfaces for timelike subregions cannot be determined solely by the boundary conditions. In the following, we demonstrate how to construct the specific RT surfaces through analytical continuation of the Euclidean counterparts. An interesting feature is that the RT surfaces naturally extend into the interior of the horizon and the complexified geometry.

Specifically, we uncover a duality for the RT surfaces: the area of the RT surface inside the horizon (red line in Fig.\ref{general}) is related to the area outside the horizon (black line in Fig.\ref{general}), as well as to the RT surfaces for spacelike regions on the Cauchy surface $t=0$ in the regions $[-t_0,t_0]$. This relation is closely tied to causality. As noted above, the RT surfaces associated with spacelike subregions lie outside the horizon, suggesting that the relation can be interpreted as a duality between the RT surfaces inside and outside the horizon. This result is remarkable. In principle, one can construct infinitely many surfaces anchored on the timelike boundary subregion, some of which may cross the horizon and even reach the singularity. Generally, one would not expect the portion inside the black hole to be determined by data outside. However, we demonstrate that certain extremal surfaces do precisely this.

Moreover, the relation between timelike and spacelike EE can be applicable to black hole physics, with an interpretation the duality between the geometric quantities inside and outside horizion. The RT surfaces inside the horizon encode information about the black hole's interior geometry. Our duality relation asserts that this information can be reconstructed using the RT surfaces outside the horizon.  This implies that an observer outside the horizon can learn the geometry inside the horizon by measuring the area of the RT surfaces outside the horizon. In the context of black hole complementarity, the information inside the black hole can be reconstructed from the exterior degrees of freedom, which means that the information in the interior of a black hole is redundant \cite{Heemskerk:2012mn, Susskind:2012uw}. The duality relation in this letter can also be interpreted as a manifestation of black hole complementarity at the level of classical spacetime.

\section{General setup for the Ryu-Takayanagi surface}

\begin{figure}[htbp]
  \centering
  \includegraphics[width=0.4\textwidth]{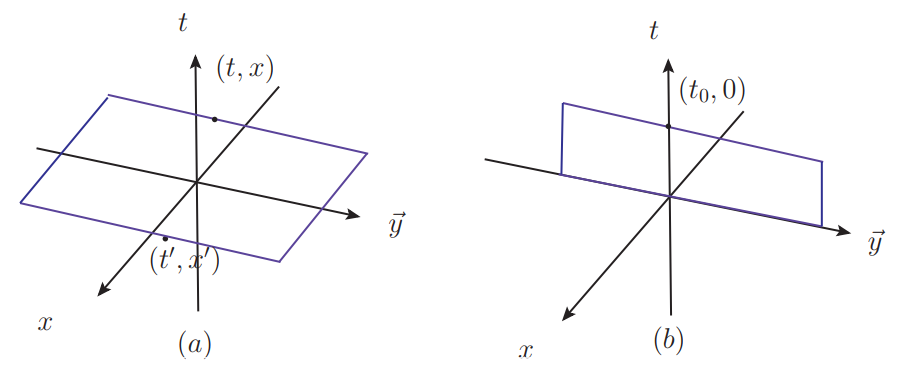}
  \caption{(a) A general timelike strip. The strip extends along the $\vec{y}$ direction, with the coordinates on $t-x$ plane $(t,x)$ and $(t',x')$ where $\vec{y}=0$. The separation between these two points are assumed timelike. (b) A special case that the strip lies along $x=0$.}
  \label{strip}
\end{figure}

Consider the asymptotically AdS$_{d+1}$ metric 
\bea\label{bulk_general}
ds^2=\frac{1}{z^2}\left(-f(z)dt^2+\frac{dz^2}{g(z)}+dx^2+d\vec{y}^2 \right),
\eea
where we set the radius of AdS $L=1$ and $d\vec{y}^2=\sum_{i=1}^{d-2}dy_i^2$. The AdS boundary is at $z = 0$.  We assume that there exists a Killing horizon $z=z_h$ with $f(z_h)=0$  for the Killing vector $\partial_t$. The spacetime is divided into regions inside and outside the horizon, as shown in Fig.\ref{general}. Generally, let us consider a timelike subregion $A_L$ to be a strip between $(t,x,\vec{y})$ and $(t',x',\vec{y}')$, denoted by $s(t,x;t',x')$, see Fig.\ref{strip}.


The procedure to construct the RT surface for $A_L$ is as follows. The Euclidean section of the solution (\ref{bulk_general}) can be obtained by $t\to -i \tau$. The Euclidean solution is defined for $z\ge z_h$. On the boundary, we will consider the subsystem $A_E$ of the strip between $(\tau,x,\vec{y})$ and $(\tau',x',\vec{y}')$, denoted by $s_E(\tau,x;\tau',x')$. First, we solve for the RT surface associated with the subsystem $A_E$. By symmetry, we can parameterize the RT surface as $\tau=\tau(z)$ and $x=x(z)$. There exists two conserved constant for the RT surface, which satisfy
\bea\label{Euclidean_condition}
&&\tau'(z)^2=\frac{p_\tau^2}{f(z)g(z)[f(z)L^2 z^{2(1-d)}-(f(z)p_x^2+p_\tau^2)]},\nn \\
&&x'(z)^2=\frac{f(z)p_x^2}{g(z)[f(z)L^2 z^{2(1-d)}-(f(z)p_x^2+p_\tau^2)]}.
\eea
In the Euclidean solution $f(z)\ge 0$, there exists a turning point of the RT surface $z=z_\tau\le z_h$ such that $\tau'(z_\tau), x'(z_\tau)\to \infty$. This leads to the condition
\bea\label{turning_point_ztau}
f(z_\tau)L^2 z_\tau^{2(1-d)}-(f(z_\tau)p_x^2+p_\tau^2)=0.
\eea
By using the conditions $\int_{0}^{z_\tau}dz \tau'(z)=\frac{\tau-\tau'}{2}$ and $\int_{0}^{z_\tau}=\frac{x-x'}{2}$, one could obtain the RT surface in the Euclidean section.

On the other hand if we consider a strip $A_L$ in the Lorentzian metric (\ref{bulk_general}), the RT surface with parameterization $t=t(z)$ and $x=x(z)$ would satisfy a  relation similar to (\ref{Euclidean_condition}),
\bea\label{Lorentzian_condition}
&&t'(z)^2=\frac{p_t^2}{f(z)g(z)[f(z)L^2 z^{2(1-d)}-(f(z)p_x^2-p_t^2)]},\nn \\
&&x'(z)^2=\frac{f(z)p_x^2}{g(z)[f(z)L^2 z^{2(1-d)}-(f(z)p_x^2-p_t^2)]}.
\eea
If the separation between $(t,x)$ and $(t',x')$ is timelike, there may not exist a real turning point $z_t$ for the RT surface. 
In general, these types of RT surface may pass through the horizon at $z=z_h$ and possibly extend to the singularity inside the horizon. However, unlike in the Euclidean case, there is no way to fix the values of $p_t$ and $p_x$ in the Lorentzian geometry. 

A specific choice is the analytical continuation of the RT surface $\tau(z)$ and $x(z)$ in the Euclidean metric. By comparing the Eq.(\ref{Euclidean_condition}) with the Wick rotation $\tau\to i t$ and (\ref{Lorentzian_condition}), the conserved constants are fixed by 
\bea\label{analytical_continuation}
&&p_t^2=-p_\tau^2|_{\tau\to i t,\tau'\to i t'},\nn \\
&&p_x^2=p_x^2|_{\tau\to i t,\tau'\to i t'}.
\eea
The turning point would also have a natural analytical result 
\bea
z_t:=z_\tau|_{\tau\to it,\tau'\to i t'}.
\eea 
With these conditions one could solve the RT surface $t(z)$ and $x(z)$. We will present some examples later. The procedure outlined above provides a method for constructing specific RT surfaces for timelike subregions. As mentioned, these RT surfaces extend into the horizon, offering a way to probe the interior of the black hole. After the analytical continuation (\ref{analytical_continuation}), there are no turning points in the Lorentzian geometry. This suggests that the RT surface should be extended into the complexified geometry, consistent with the framework proposed in \cite{Heller:2024whi}.

The area of the RT surface is given by
\bea\label{area_term}
&&\mathcal{A}(t,x;t',x')=2\int_{C}dz \mathcal{L},\;\nn\\
&&\mathcal{L}=\frac{R^{d-2}}{z^{2(d-1)}}\frac{\sqrt{f(z)}}{\sqrt{g(z)[f(z)L^2 z^{2(1-d)}-(f(z)p_x^2-p_t^2)]}},
\eea
where $R$ is the IR cut-off for the $y_i$ directions, $C$ is a path on the complex $z$ plane connecting the boundary $z=\delta$ to the turning point $z_t$, which is generally a branch point of the integrand in (\ref{area_term}).  
\begin{figure}[htbp]
  \centering
  \includegraphics[width=0.35\textwidth]{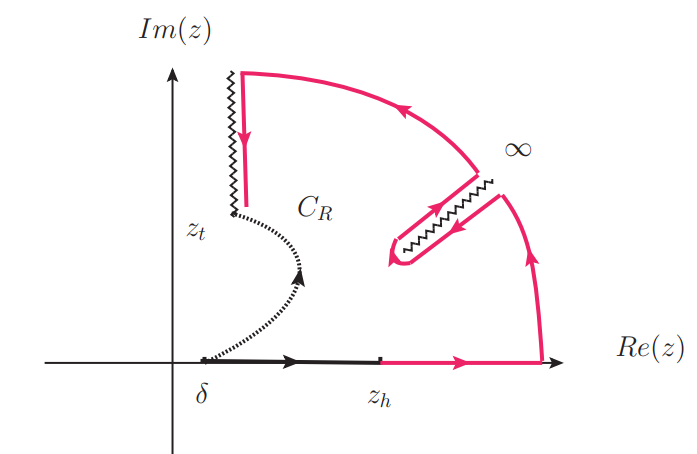}
  \caption{ The evaluation of the integral (\ref{area_term}) requires specifying a path $C$ in the complex $z$-plane that connects $z=\delta$ to $z=z_t$. We assume that the path $C_R$ (dotted line) corresponds to the correct RT surface. For our purpose, however, we choose a path that first runs from $z=\delta$ to $z=z_h$ (solid black line), and then continues from $z_h$ to $z_t$ (solid red line).}
  \label{path_fig}
\end{figure}
To evaluate the integral (\ref{area_term}), one must specify the path $C$. However, our method does not directly determine this choice. In principle, we expect the RT surface for the strip to be unique, corresponding to the path $C_R$ on the complex-$z$ plane. Since (\ref{area_term}) is a contour integral, the result remains unchanged for different paths as long as the regions enclosed by them contain no singularities. Therefore, one may equivalently choose the path $C$ shown in Fig.~\ref{path_fig}, in which case the associated bulk surface extends across the black hole horizon and reaches the singularity.

We can thus define and evaluate the surface areas on both sides of the horizon:
\bea\label{definition_in_out}
&&\mathcal{A}_{\text{out}}=2\int_{\delta}^{z_h}dz\mathcal{L},\nn \\
&&\mathcal{A}_{\text{in}}=2\int_{z_h}^{z_t}dz\mathcal{L}.
\eea
By definition we have $\mathcal{A}(t,x;t',x')=\mathcal{A}_{\text{in}}+\mathcal{A}_{\text{out}}$. 

In the case of higher-dimensional black holes, analytical results can only be obtained via perturbative methods. Alternatively, numerical methods can be used to determine turning points. Let us focus on the strip $s(0,0;t_0,0)$. For this setup, we solve Eq.(\ref{Lorentzian_condition}) with $p_x=0$. The key issue is to determine the turning point $z_t$, which is generally complex-valued. By symmetry, we expect the boundary conditions for Eq.(\ref{Lorentzian_condition}) to be $t(0)=0$ and $t(z_t)=0$. Consequently, we can numerically solve the differential equation in a manner similar to the spacelike case, except that both $z_t$ and $\mathcal{A}$ are complex-valued. In the following, we present numerical results for the higher-dimensional black hole examples.

\section{The duality}
The RT surfaces we construct provide a holographic realization of timelike EE for the strip subregions.In the QFTs, the timelike EE can be evaluated by analytically continuing correlators involving twist operators. Our construction of RT surfaces for timelike subregions thus serves as the bulk dual of this analytic continuation. Consider the specific case $A_L=s(0,0;t_0,0)$, as illustrated in Fig.\ref{strip}(b).  The timelike EE is associated with the correlator of twist operator $\Sigma_n$,  which are generally non-local. Specifically, $S_E\sim \langle \Sigma_n(t_0,0)\Sigma_n(0,0)\rangle_{\text{QFT}^n}$, where the subscript denotes $n$-copies of the theory. By causality, the operators $\Sigma_n(t_0,0)$ can be decomposed as operators on the Cauchy surface $t=0$ within the region $x\in[-t_0,t_0]$, that is the strip $s(0,-t_0;0,t_0)$. Consequently, $S_E$ is expected to be represented by correlators on the Cauchy surface, which may also exhibit bulk geometric duality. In the following examples, we demonstrate that only  $\mathcal{A}(0,x;0,x')$ and $\partial_t \mathcal{A}(0,x;0,x')$ are required, both of which depend solely on data outside the black hole horizon. For more general states, causality implies that the timelike EE should still depend only on the expectation values of operators localized in the spacelike strip $s_L(0,-t_0;0,t_0)$. However, in such cases, additional contributions from other operators in the theory are also expected to appear \cite{Guo:2025mwp}. According to the subregion/subregion duality, or equivalently the entanglement wedge reconstruction theorem \cite{Czech:2012bh,Jafferis:2015del,Dong:2016eik} in AdS/CFT, these new contributions can still be mapped to the bulk entanglement wedge associated with $s(0,-t_0;0,t_0)$, which lies outside the horizon. In summary, we expect that the bulk surface encoding the timelike EE can be reconstructed entirely from information outside the horizon. Thus, the duality relation remains valid for general states, although the resulting expressions may be too complicated to be practically useful.

\textit{BTZ black hole and AdS-Rindler}\ The metric of BTZ black hole is given by 
\bea\label{metric_BTZ}
ds^2=\frac{1}{z^2}\left[-f(z)dt^2 +\frac{dz^2}{f(z)}+dx^2\right],
\eea
with $f(z)=1-\frac{z^2}{z_h^2}$, where $z=z_h$ is the black hole horizon. The singularity is at $z=\infty$. We consider the timelike interval $[0,t_0]$ on $x=0$. For the strip $s_E(0,0;\tau_0,0)$ we can obtain $p_{\tau_0}=\frac{\sin\frac{\tau_0}{z_h}}{z_h(1-\cos\frac{\tau}{z_h})}$. By using (\ref{analytical_continuation}) we have
\bea
p_{t_0}^2=\frac{1}{z_h^2}\left(\coth\frac{t_0}{2z_h}\right)^2.
\eea
The turning point is given by $z_{t_0}=i z_h\sinh\frac{t_0}{2z_h}$. The length of RT surface inside and outside the horizon is given by
\bea
&&\mathcal{A}_{\text{in}}=2\log \left[(1+\coth\frac{t_0}{2z_h})\sinh\frac{t_0}{2z_h} \right]+i\pi,\nn \\
&&\mathcal{A}_{\text{out}}=2\log\left[\frac{2z_h}{\delta}\frac{1}{1+\coth\frac{t_0}{2z_h}} \right].
\eea
See the Supplemental Material for calculation details. The above result yields the expected value for the timelike EE of the interval $[0,t_0]$.
It can be shown that the RT surfaces inside and outside the horizon satisfy the relation
\bea\label{AdS3_relation}
&&\mathcal{A}_{\text{in}}=\frac{1}{2}\left(\mathcal{A}(0,-t_0;0,0)+\mathcal{A}(0,0;0,t_0)\right)\nn \\
&&\phantom{\mathcal{A}_{\text{in}}=}+\frac{1}{2}\int_{-t_0}^{t_0}dx\partial_t \mathcal{A}(0,x;0,0)-\mathcal{A}_{\text{out}}.
\eea
The detailed derivation is presented in the Supplemental Material.
For $z_h=1$, the metric corresponds to AdS-Rindler coordinates, which cover only a portion of AdS. 

~\\
\textit{Higher dimensional examples}\ Let us firstly consider the pure AdS$_{d+1}$ ($d\ge 2$). The metric is given by (\ref{bulk_general}) with $f(z)=g(z)=1$. There exists a horizon at $z=\infty$. We consider the strip $s(0,0;t_0,0)$. Using the same procedure, we can evaluate the RT surface in the Euclidean metric for the subsystem $s_E(0,0;\tau_0,0)$. We find
\bea
p_\tau^2=z_\tau^{2-2d},\quad \text{with}\quad z_{\tau}=\frac{\tau_0\Gamma(\frac{1}{2(d-1)})}{2\sqrt{\pi}\Gamma(\frac{d}{2(d-1)})}.
\eea
With the analytical continuation $\tau_0\to i t_0$, we obtain
\bea
p_t^2=i^{2(2-d)}\left(\frac{t_0\Gamma(\frac{1}{2(d-1)})}{2\sqrt{\pi}\Gamma(\frac{d}{2(d-1)})}\right)^{2-2d}.
\eea
The turning point can also be obtained by analytical continuation $z_t=\frac{it_0\Gamma(\frac{1}{2(d-1)})}{2\sqrt{\pi}\Gamma(\frac{d}{2(d-1)})}$. 
The area of the RT surface is given by
\bea\label{timelike_strip}
\mathcal{A}(0,0;t_0,0)=\frac{2R^{d-1}}{(d-2)\delta^{d-2}}+2\kappa_d \frac{R^{d-1}(-i)^{d}}{t_0^{d-2}},
\eea
with $\kappa_d=\frac{\pi^{\frac{d-1}{2}} 2^{d-2}}{d-2}(\frac{\Gamma(\frac{d}{2(d-1)})}{\Gamma(\frac{1}{2(d-1)})})^{d-1}$. See the supplemental material for calculation details. The results are consistent with \cite{Heller:2024whi, Xu:2024yvf}. For the spacelike strip, the first-order temporal derivative satisfies the following relation \cite{Xu:2024yvf},
\bea\label{first_derivative_strip_vacuum}
x^{d-2}\partial_t \mathcal{A}(0,x;0,0)= 2i R^{d-1}\kappa_d  (d-2)\pi \delta(x).
\eea
Just like the AdS$_3$ case, the RT surface approaches the horizon $z=\infty$ and extends into the complexified geometry. The areas of the RT surface inside and outside the horizon can be evaluated. Guided by Eq. (\ref{first_derivative_strip_vacuum}) and inspired by the AdS$_3$ relation (\ref{AdS3_relation}), we construct the following equation
\bea\label{duality vacuum}
&&\mathcal{A}(0,0;t_0,0)\nn \\
&&=\frac{1}{2}\left(\mathcal{A}(0,-t_0;0,0)+\mathcal{A}(0,0;0,t_0) \right)\nn \\
&&+\frac{i[(-i)^{d-2}-1]}{t_0^{d-2}(d-2)\pi} \int_{-t_0}^{t_0}dx x^{d-2}\partial_t \mathcal{A}(0,x;0,0).
\eea
See the Supplemental Material for the derivation.
One can directly verify the above formula using Eqs. (\ref{timelike_strip}) and (\ref{first_derivative_strip_vacuum}). Notably, in the limit $d\to 2$, the formula reduces to the AdS$_3$ case given by Eq. (\ref{AdS3_relation}).

Now, let us consider the black hole background with $f(z)=g(z)=1-\frac{z^d}{z_h^d}$. We will still consider the timelike strip $s(0,0;t_0,0)$. In this case, obtaining an analytical expression for the RT surface area is challenging. However, for small strips satisfying $t_0\ll z_h$, the problem can be solved perturbatively. In this section, we set $z_h=1$ and assume $t_0\ll 1$, keeping only the leading-order contributions.

Both the timelike and spacelike RT surfaces receive thermal corrections. Let $\mathcal{A}_{bh}(0,0;0,x)$ denote the area of the RT surface for the spacelike strip with width $x$ on the Cauchy surface $t=0$. For $x\ll 1$, the leading-order thermal correction is given by 
\bea
&&\mathcal{A}_{bh}(0,0;0,x)=\mathcal{A}(0,0;0,x)+\delta \mathcal{A}_{bh}(0,0;0,x),\nn\\
&& \text{with}\ \delta \mathcal{A}_{bh}(0,0;0,x)=\alpha_d x^2+O(x^{2+d}),
\eea
where $\delta \mathcal{A}_{bh}$ represents the thermal correction to the EE of the spacelike strip. The results for $\alpha_d$ are provided in the Supplemental Material. 

For the timelike strip $s(0,0;t_0,0)$ we can also obtain the thermal correction. The procedure is the same as before. We consider the strip $s_E(0,0;\tau_0,0)$ in the Euclidean spacetime. By some calculations, we obtain the turning point $z_{\tau_0}$. The explicit results are provided in the Supplemental Material.
After performing the analytical continuation $\tau\to i t$, we obtain the turning point $z_{t_0}=z_{\tau_0}|_{\tau_0\to i t_0}$. We consider the RT surface connecting $z=\delta$ to $z_{t_0}$, with the path $C$ on the complex $z$-plane chosen similarly to the vacuum case. Given this path, we can then evaluate the area of the RT surface.
\bea
\mathcal{A}_{bh}(0,0;t_0,0)=\mathcal{A}(0,0;t_0,0)+\beta_d t_0^2+O(t_0^{2+d}),
\eea
where $\beta_d$ is a constant that depends only on $d$. The explicit results are provided in the Supplemental Material.

\begin{figure}[htbp]
  \centering
  \includegraphics[width=0.5\textwidth]{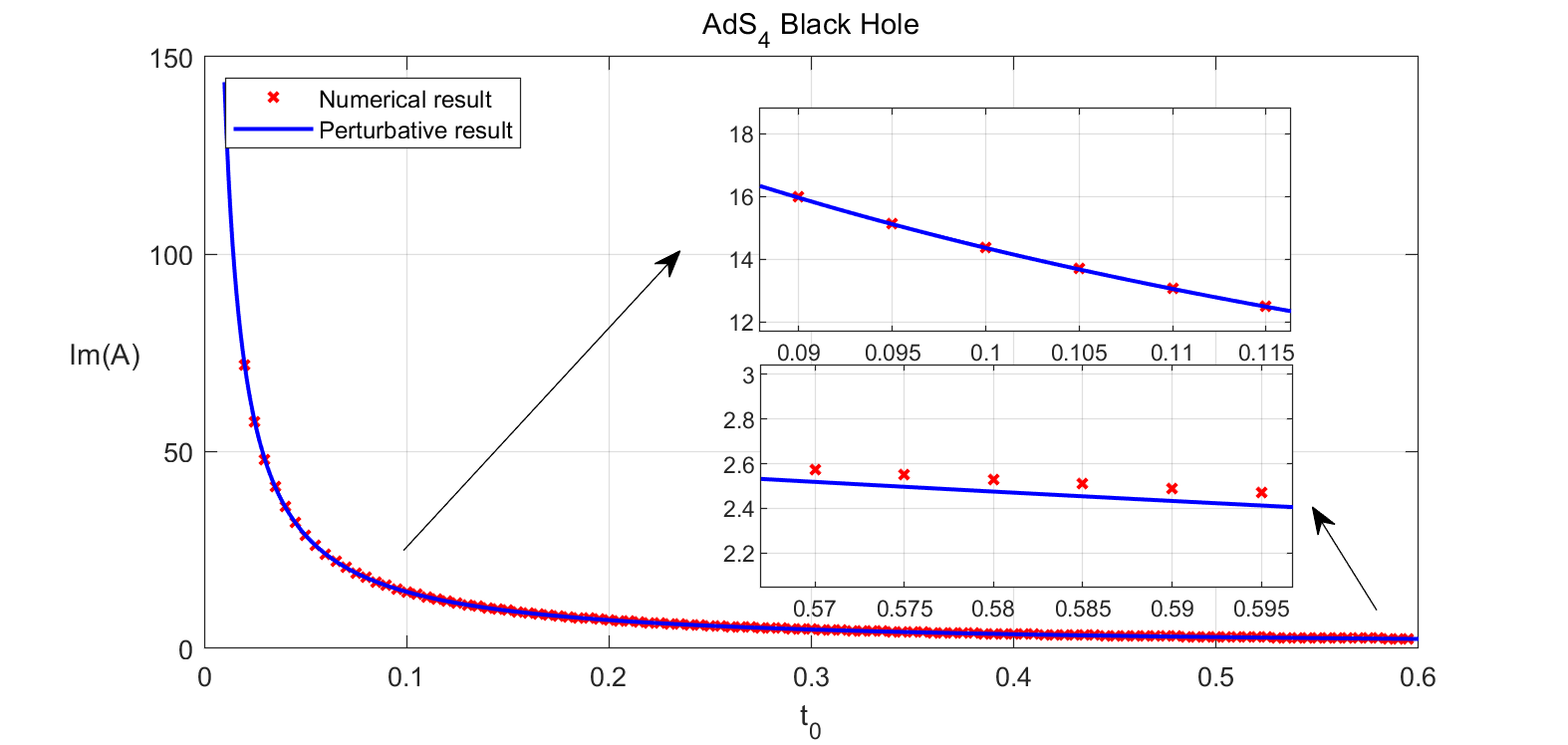}
  \caption{Plot of the imaginary part of the RT surface area $\mathcal{A}(0,0;t_0,0)$ versus interval length $t_0$ for AdS$_4$ black hole. The red crosses denote numerical results, while the blue curve represents the perturbative analytical solution. We set $z_h=R=1$. The interval length $t_0$ is sampled over $(0,0.6)$ with spacing 0.005.}\label{fig_AdS4}
\end{figure}

\begin{figure}[htbp]
  \centering
  \includegraphics[width=0.5\textwidth]{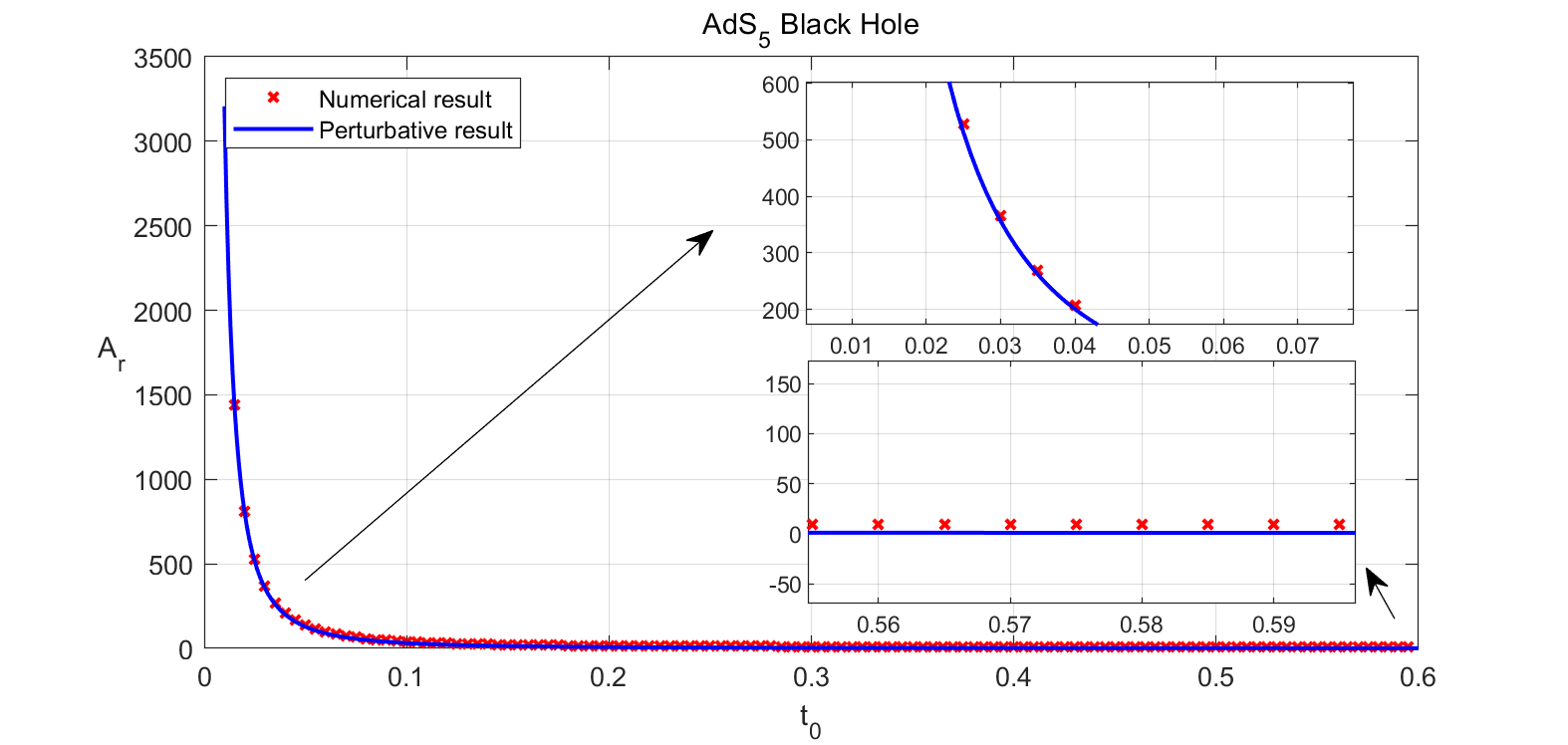}
  \caption{Plot of the regularized RT surface area $\mathcal{A}_r=\mathcal{A}_{bh}(0,0;t_0,0)-\frac{2}{3\delta^{3}}$ versus interval length $t_0$ for AdS$_5$ black hole. The red crosses denote numerical results, while the blue curve represents the perturbative calculation. Here, we set $z_h=R=1$. The interval length $t_0$ is sampled over $(0,0.6)$ with spacing 0.005.}\label{fig_AdS5}
\end{figure}

For higher-dimensional black holes, the area of the RT surface can also be obtained numerically. In Fig.\ref{fig_AdS4} (AdS$_4$ black hole) and Fig.\ref{fig_AdS5} (AdS$_5$ black hole), we compare the numerical results with those from the perturbative approach. The agreement holds at small $t_0\ll z_h$, with discrepancies appearing at $t_0\sim z_h$.

We now demonstrate that the relation (\ref{first_derivative_strip_vacuum}) remains valid in the black hole background. Consider a strip $s(t,x;0,0)$ with $t\ll 1$ while ensuring $-t^2+x^2>0$, i.e., a spacelike strip.  The holographic EE can be computed perturbatively in this regime. Due to time-reversal symmetry, thermal corrections to the RT surface $\mathcal{A}_{bh}(t,x;0,0)$ cannot introduce terms linear in $t$, the leading-order corrections must be at most $O(t^2)$. Consequently, we find $\partial_t\mathcal{A}_{bh}(t,x;0,0)|_{t\to 0}=\partial_t\mathcal{A}(t,x;0,0)|_{t\to 0}$, implying that thermal corrections do not affect the first-order temporal derivative of the spacelike RT surface.

Notably, the corrections to the spacelike and timelike strips differ, i.e., $\alpha_d\ne \beta_d$. However, we find that they obey the universal relation
\bea
\frac{\beta_d}{\alpha_d}=-\frac{d-3}{d-1}.
\eea
See the Supplemental Material for more details. Thus, the relation (\ref{duality vacuum}) can be modified as
\bea\label{duality_blackhole}
&&\mathcal{A}_{bh}(0,0;t_0,0)\nn \\
&&=\frac{1}{2}\left(\mathcal{A}_{bh}(0,-t_0;0,0)+\mathcal{A}_{bh}(0,0;0,t_0)  \right)\nn \\
&&-\frac{d-2}{d-1}(\delta\mathcal{A}_{bh}(0,-t_0;0,0)+\delta\mathcal{A}_{bh}(0,0;0,t_0))\nn \\
&&+\frac{i[(-i)^{d-2}-1}{t_0^{d-2}(d-2)\pi} \int_{-t_0}^{t_0}dx x^{d-2}\partial_t \mathcal{A}_{bh}(0,x;0,0).
\eea
The derivation is presented in the Supplemental Material.
The right-hand side of the above equation only includes information outside the black hole horizon. For $d=2$ there are no thermal corrections on the right-hand side, which is consistent with formula (\ref{AdS3_relation}) for the BTZ black hole. We can divide the RT surface for the timelike strip into the inside and outside parts of the horizon. Thus, the RT surface inside the horizon can be constructed using quantities outside the horizon.

\section{Discussion}
In this letter, we demonstrate how to construct the RT surface for the timelike strip via analytical continuation. We also uncover a duality between the RT surfaces inside and outside the black hole horizon, which can be interpreted as a realization of the concept of black hole complementarity: the information inside the horizon can be reconstructed from the degrees of freedom outside the horizon.

In our construction, it seems necessary to consider the RT surface in the complexified geometry. The role of the complexified geometry in understanding the interior of black holes remains unclear. 
If matter is included in the bulk,  we would generally have $f\ne g$. It has been shown in \cite{Frenkel:2020ysx} that the metric near the singularity would resemble a more general Kasner universe \cite{Kasner:1921zz,Belinski:1973zz}. It would be interesting to investigate whether it is possible to construct the duality relation in this case and explore the properties of the singularity through this duality. Additionally, it would be interesting to investigate whether a similar duality persists in a black hole with Hawking radiation \cite{Penington:2019npb,Almheiri:2019psf,Almheiri:2019hni,Almheiri:2020cfm}.

~\\

{\bf Acknowledgements}
We would like to thank Song He, Yan Liu, Rong-Xin Miao, Tadashi Takayanagi, Run-qiu Yang, Jiaju Zhang and Yu-Xuan Zhang for useful discussions. WZG would also like to express my gratitude to the Gauge/Gravity Duality 2024 conference, where part of the content of this work was presented.
WZG is supposed by the National Natural Science Foundation of China under Grant No.12005070 and the Hubei Provincial Natural Science Foundation of China under Grant No.2025AFB557.

\cleardoublepage\onecolumngrid
\clearpage
\begin{center}
{\large\bfseries Supplementary Material}
\end{center}
\vspace{1em}
\appendix
We now present some useful formulae. Recall the equations for the RT surface in the Euclidean solution,
\bea\label{Euclidean_condition}
&&\tau'(z)^2=\frac{p_\tau^2}{f(z)g(z)[f(z)L^2 z^{2(1-d)}-(f(z)p_x^2+p_\tau^2)]},\nn \\
&&x'(z)^2=\frac{f(z)p_x^2}{g(z)[f(z)L^2 z^{2(1-d)}-(f(z)p_x^2+p_\tau^2)]}.
\eea
The equations for the RT surface in Lorentzian solution are
\bea\label{Lorentzian_condition}
&&t'(z)^2=\frac{p_t^2}{f(z)g(z)[f(z)L^2 z^{2(1-d)}-(f(z)p_x^2-p_t^2)]},\nn \\
&&x'(z)^2=\frac{f(z)p_x^2}{g(z)[f(z)L^2 z^{2(1-d)}-(f(z)p_x^2-p_t^2)]}.
\eea
The area of the RT surface is given by
\bea\label{area_term}
&&\mathcal{A}(t,x;t',x')=2\int_{C}dz \mathcal{L},\;\nn\\
&&\mathcal{L}:=\frac{R^{d-2}}{z^{2(d-1)}}\frac{\sqrt{f(z)}}{\sqrt{g(z)[f(z)L^2 z^{2(1-d)}-(f(z)p_x^2-p_t^2)]}},
\eea
where $R$ is the IR cut-off for the $y_i$ directions, $C$ is a path on the complex $z$ plane connecting the boundary $z=\delta$ to the turning point $z_t$. In the following discussion, we will set $R=1$.

\section{I: Examples in AdS$_3$}\label{BTZ_section}

 The simplest example is pure AdS$_3$ in Poincare coordinate with metric $ds^2=\frac{-dt^2+dz^2+dx^2}{z^2}$.  Let us construct the timelike RT surface associated with the interval $[0,t_0]$ on the slice $x=0$. Performing the procedure in the previous section, we can obtain $p_{\tau_0}=\frac{2}{\tau_0}$. By analytical continuation, we have
\bea
p_{t_0}^2=\frac{4}{t_0^2}.
\eea
One could solve for the RT surface in the Lorentzian metric and obtain $z_t= i\frac{t_0}{2}$. There exists a horizon at infinity $z=+\infty$ in the Poincar\'e coordinate. Through some calculations, the area of the RT surface inside the horizon is given by $\mathcal{A}_{\text{in}}=i\pi$ and $\mathcal{A}_{\text{out}}=2\log\frac{t_0}{\delta}$. By definition, the total length of the RT surface is $\mathcal{A}(0,0;t_0,0)=\mathcal{A}_{\text{in}}+\mathcal{A}_{\text{out}}$.  It is straightforward to show that the following relation holds,
\bea\label{AdS3_relation}
&&\mathcal{A}_{\text{in}}=\frac{1}{2}\left(\mathcal{A}(0,-t_0;0,0)+\mathcal{A}(0,0;0,t_0)\right)\nn \\
&&\phantom{\mathcal{A}_{\text{in}}=}+\frac{1}{2}\int_{-t_0}^{t_0}dx\partial_t \mathcal{A}(0,x;0,0)-\mathcal{A}_{\text{out}}.
\eea
By RT formula $S=\frac{\mathcal{A}(0,0;t_0,0)}{4G}$, we obtain the expected result for the timelike EE of the interval $[0,t_0]$. The relation (\ref{AdS3_relation}) is precisely the timelike and spacelike EE relation constructed in \cite{Guo:2024lrr} for the CFT vacuum state. Notably, the right-hand side of (\ref{AdS3_relation}) involves only the RT surfaces outside the horizon ($z=\infty$). 
\\

It is also straightforward to calculate the RT surfaces in the BTZ black hole and the AdS-Rindler wedge. The metric of the 3-dimensional AdS-Rindler wedge can be seen as the case $z_h=1$. The AdS-Rindler coordinate is generally written as
\bea
ds^2=-(r^2-1)dt^2+\frac{d r^2}{r^2-1}+r^2 dx^2,
\eea
$1<r<+\infty$, $-\infty<x<+\infty$ and $-\infty<t<+\infty$. This coordinate system covers only part of the global coordinates. A horizon exists at $z=1$. By performing the coordinate transformation $r=\frac{1}{z}$, the metric becomes
\bea
ds^2=\frac{1}{z^2}\left[-f(z)dt^2 +\frac{dz^2}{f(z)}+dx^2\right],
\eea
where $f(z)=1-\frac{1}{z^2}$. 

Consider the Euclidean section of the metric of BTZ, 
\bea\label{Euclidean_Rindler}
ds^2=\frac{1}{z^2}\left[f(z)d\tau^2 +\frac{dz^2}{f(z)}+dx^2\right],
\eea
where $\tau$ is the Euclidean time with $\tau\sim \tau+2\pi z_h$. The boundary CFT is thus in the thermal state with an inverse temperature $\beta=2\pi  z_h$.

As discussed in the main text, our focus is on the RT surface associated with the timelike interval $[0,t_0]$ and the temporal derivative of the RT surface for the spacelike interval $[x_0,x_1]$. The results presented above pertain to the RT surface corresponding to the spacetime interval between the points $(t,x)$ and $(t',x')$. Our approach involves evaluating these results in the Euclidean metric (\ref{Euclidean_Rindler}), after which the necessary results are obtained through the analytical continuation $\tau\to i t$.

Let us consider an interval $A_E$ between $(\tau,x)$ and $(\tau',x')$ on the boundary of Euclidean BTZ or AdS-Rindler (\ref{Euclidean_Rindler}). We would like to evaluate the RT surface for $A_E$, which can be parameterized as $\tau=\tau(z)$ and $x=x(z)$. There exists two conserved constants for the RT surface, 
\bea
&&p_\tau=\frac{ f(z)^{3/2}\tau'(z)}{z\sqrt{1+f(z) x'(z)+f(z) \tau'(z)^2}},\nn \\
&&p_x=\frac{ f(z)^{1/2} x'(r)}{z\sqrt{1+f(z) x'(z)+f(z) \tau'(z)^2}}.
\eea 
 There exists turning point of the geodesic with $z=z_*$ satisfying $\tau'(z_*)=\infty$ and $x'(z_*)=\infty$. We also have the relation
\bea
-f(z_*)L^2+f(z_*)p_x^2 z_*^2+p_\tau^2 z_*^2=0,
\eea
and the conditions
\bea
\int_{0}^{z_*}dz \tau'(z)dz=\frac{\tau-\tau'}{2},\quad  \int_{0}^{z_*}dz x'(z)dz =\frac{x-x'}{2}.\;
\eea
With these conditions we can obtain
\bea
&&p_\tau=\frac{\sin\frac{\tau-\tau'}{z_h}}{\cosh\frac{x-x'}{z_h}-\cos\frac{\tau-\tau'}{z_h}},\nn\\
&&p_x=\frac{\sinh\frac{x-x'}{z_h}}{\cosh\frac{x-x'}{z_h}-\cos\frac{\tau-\tau'}{z_h}},\nn \\
&&z_*=z_h\sqrt{\frac{\cosh\frac{x-x'}{z_h}-\cos\frac{\tau-\tau'}{z_h}}{1+\cosh\frac{x-x'}{z_h}}}.
\eea
The length of the geodesic line in the Euclidean section is given by
\bea
\mathcal{L}=2z_h \log \left(\frac{4 z_h^2 \sinh \left(\frac{\bar w}{2 z_h}\right) \sinh \left(\frac{w}{2 z_h}\right)}{\delta^2}\right),
\eea
where $w=x+i\tau$ and $\bar w=x-i \tau$.
Taking $z_h=\frac{\beta}{2\pi}$ into the above equation and using the RT formula, we obtain the expected result for EE in thermal state with inverse temperature $\beta$. One could obtain the results for the timelike regions by using the analytical continuation $\tau \to it$.

\begin{figure}[htbp]
  \centering
  \includegraphics[width=0.4\textwidth]{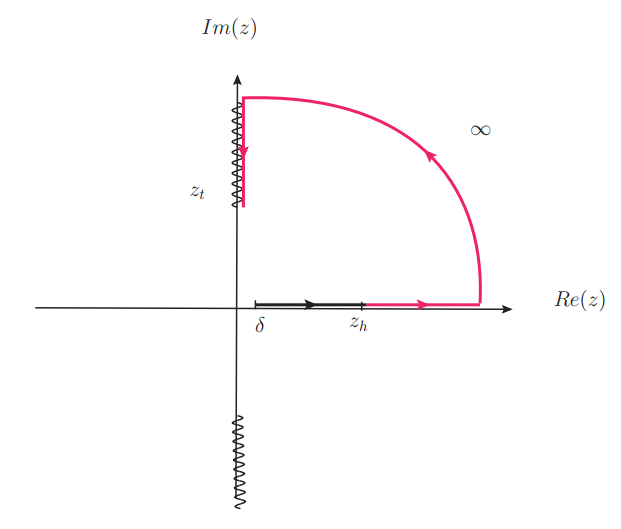}
  \caption{The path chosen for the integration is given by (\ref{BTZintegration}).}
  \label{BTZpath}
\end{figure}

We can also obtain the same results by following the procedure discussed in the main text. Consider the interval on the time coordinate $[0,t_0]$. One could obtain $p_t$, $p_x$ and $z_t$ through analytical continuation. The results are
\bea
p_t=\frac{\coth\frac{t_0}{2z_h}}{z_h},\quad p_x=0,\quad z_t=i z_h\sinh\frac{t_0}{2z_h}.
\eea
With these, one can solve for the timelike RT curve $t=t(z)$. The length of geodesic line is given by
\bea\label{BTZintegration}
\mathcal{L}=2\int_{C} dz \frac{z_h}{z}\frac{1}{\sqrt{p_t^2z^2z_h^2-(z^2-z_h^2)}},
\eea 
where the path $C$ is chosen as shown in Fig.\ref{BTZpath}.

It can be shown directly that the integration yields the correct results for the timelike EE. It is also straightforward to calculate $\mathcal{A}_{\text{in}}$ and $\mathcal{A}_{\text{out}}$ by the definition.


\section{II: Higher-Dimensional Strip in the Vacuum State}

In this section, we present the details of the calculation for the higher-dimensional strip. For the vacuum state, where $f=g=1$, analytical results can be obtained.
We will primarily focus on $d=3$ and $d=4$, highlighting the significant differences between the even- and odd-dimensional cases. 

Consider the strip $s_E(0,0;\tau_0,0)$. By using the Eq.(\ref{Euclidean_condition}) and the conditions that $\int_0^{z_\tau}\tau'(z)dz=\frac{\tau_0}{2}$, where $z_\tau$ is the turning point of the RT surface. One could obtain
\bea
p^2_\tau=z_\tau^{2-2d},
\eea
with 
\bea
z_\tau=\frac{\tau_0 \Gamma(\frac{1}{2(d-1)})}{2\sqrt{\pi}\Gamma(\frac{d}{2(d-1)})}.
\eea
With the continuation $\tau\to it$, we find
\bea
p_t^2=(-1)^{1-d}\left(\frac{t_0\Gamma(\frac{1}{2(d-1)})}{2\sqrt{\pi}\Gamma(\frac{d}{2(d-1)})}\right)^{2-2d}.
\eea
Note that $p_t$ is real for even $d$, while it is imaginary for odd $d$. This leads to a significantly different behavior in odd and even dimensions. The turning point $z_t$ can also be obtained through analytical continuation $z_t=z_\tau|_{\tau \to i t}$,
\bea
z_t=\frac{i t_0 \Gamma(\frac{1}{2(d-1)})}{2\sqrt{\pi}\Gamma(\frac{d}{2(d-1)})}.
\eea
With these results one could evaluate the area for the RT surface.
From (\ref{area_term}) we will have
\bea\label{area_function}
\mathcal{A}=\int_{C}dz \mathcal{L} \ \text{with}\ \frac{1}{z^{d-1}\sqrt{1+p_t^2z^{2(d-1)}}},
\eea
where $C$ is the path connecting $z=\delta$ to $z_t$.

On the complex $z$-plane the integrand in (\ref{area_function}) will have pole at $z=0$ and branch points 
\bea
z_{b,k}=z_t e^{\frac{2\pi i k}{2(d-1)}},\quad \text{with}\quad k=0,...,2d-3.
\eea

\begin{figure}[htbp]
  \centering
  \includegraphics[width=0.4\textwidth]{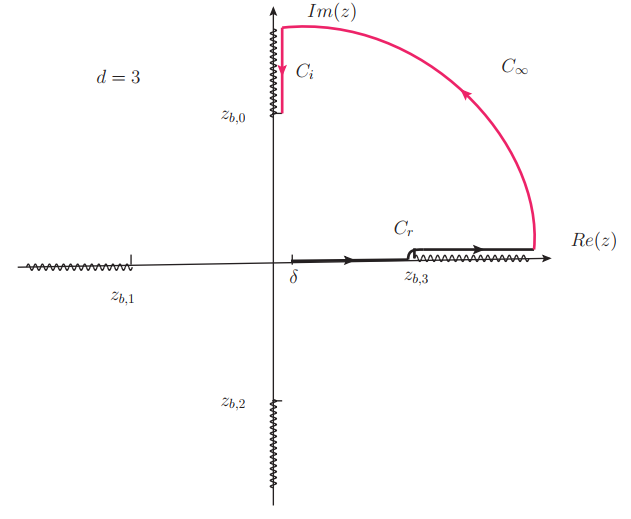}
  \caption{The path chosen for the integration is given by (\ref{area_function}) with $d=3$.}
  \label{pathd3}
\end{figure}

\begin{figure}[htbp]
  \centering
  \includegraphics[width=0.4\textwidth]{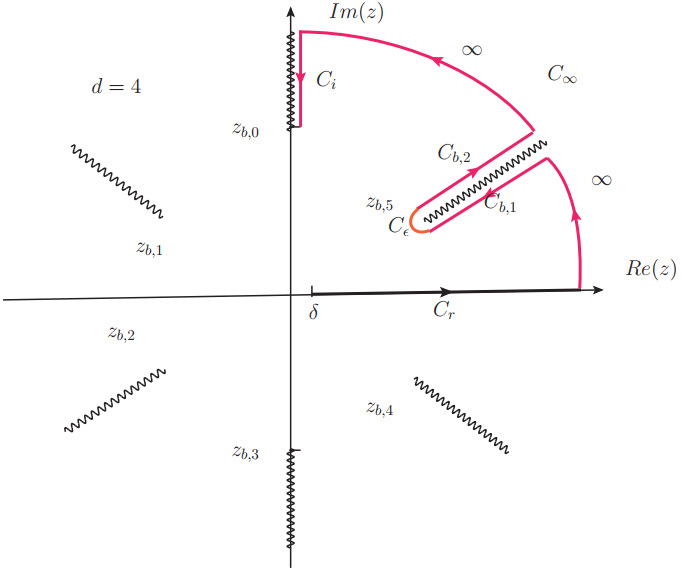}
  \caption{The path chosen for the integration is given by (\ref{area_function}) with $d=4$. }
  \label{pathd4}
\end{figure}
For $d=3$ and $d=4$, the branch cuts are shown in Fig. \ref{pathd3} and Fig. \ref{pathd4}. For $d=3$ a branch cut exists along the $Re(z)$ coordinate, whereas no such cut appears for $d=4$. We can now evaluate the area of the RT surface by carefully selecting the integration path $C:\delta\to z_t$. Our goal is to construct an RT surface that extends to the horizon, which, in Poincaré coordinates, is located at $z=\infty$. The chosen paths are illustrated in Fig. \ref{pathd3} and Fig. \ref{pathd4}.

Let us first consider the case of $d=3$. It can be easily shown that the contribution from the path $C_\infty$ is zero. After some calculations, we obtain
\bea
&&\int_{C_r}dz \mathcal{L}=\frac{1}{\delta }-\frac{2 \pi  \Gamma \left(\frac{3}{4}\right)^2}{t_0 \Gamma \left(\frac{1}{4}\right)^2}+\frac{2 i \pi  \Gamma \left(\frac{3}{4}\right)^2}{t_0 \Gamma \left(\frac{1}{4}\right)^2},\nn \\
&&\int_{C_i}dz \mathcal{L}=\frac{2 \pi  \Gamma \left(\frac{3}{4}\right)^2}{t_0 \Gamma \left(\frac{1}{4}\right)^2},
\eea
where $\delta$ is the UV cut-off. Thus we obtain the area of the RT surface
\bea
\mathcal{A}_{d=3}=\frac{1}{\delta }+\frac{2 i \pi  \Gamma \left(\frac{3}{4}\right)^2}{t_0 \Gamma \left(\frac{1}{4}\right)^2}.
\eea

For the $d=4$ case, it is also easy to show that the contributions from $C_\infty$ vanish. The contributions from the remaining paths are given by
\bea
&&\int_{C_r}dz\mathcal{L}=\frac{1}{2 \delta^2}+\frac{\sqrt{\pi } \Gamma \left(-\frac{1}{3}\right) \Gamma \left(\frac{11}{6}\right) \Gamma \left(\frac{2}{3}\right)^2}{45 t_0^2 \Gamma \left(\frac{7}{6}\right)^2},\nn \\
&&\int_{C_{b,1}+C_{b,2}}dz\mathcal{L}=\frac{i \left(\sqrt{3}-3 i\right) \pi ^{3/2} \Gamma \left(\frac{2}{3}\right)^2 \Gamma \left(\frac{5}{3}\right)}{4 t_0^2 \Gamma \left(\frac{1}{6}\right)^2 \Gamma \left(\frac{7}{6}\right)},\nn \\
&&\int_{C_i}dz\mathcal{L}=-\frac{i \pi ^2 \Gamma \left(\frac{5}{6}\right) \Gamma \left(\frac{5}{3}\right)}{2^{1/3} t_0^2 \Gamma \left(\frac{1}{6}\right)^2 \Gamma \left(\frac{7}{6}\right)}
\eea
Thus the area of the RT surface in the $d=4$ case is given by
\bea
\mathcal{A}_{d=4}=\frac{1}{2 \delta^2}+\frac{2 \pi ^{3/2} \Gamma \left(\frac{2}{3}\right)^3}{t_0^2 \Gamma \left(\frac{1}{6}\right)^3}.
\eea
By our definition we can take $\int_{C_r}dz \mathcal{L}$ as the length of the RT surface outside the horizon. The total area of the RT surface is consistent with the results expected to timelike strip, discussed in \cite{Heller:2024whi, Xu:2024yvf}. 

In the vacuum case, the RT surface $t=t(z)$ can be solved analytically.  By (\ref{Lorentzian_condition}) we have
\bea
t(z)=\int_{0}^zdz t'(z),\  \text{with}\ t'(z)=\frac{\pm i p_t}{\sqrt{z^{2(1-d)}+p_t^2}}.
\eea
The integration can be obtained analytically,
\bea
t(z)=t_0\pm \frac{i}{d}\frac{z^d}{z_t^{d-1}}~_2F_1\left[\frac{1}{2},\frac{d}{2(d-1)},\frac{3d-2}{2(d-1)},\left(\frac{z}{z_t}\right)^{2(d-1)}\right], \nn
\eea
In general, the RT surface extends into the complexified geometry, even for real values of $z$. 

For $d=3$ with $z\in (\delta,\infty)$, $t(z)$ starts from the boundary $z=\delta$. In the region $z\in (\delta,z_{b,3})$ the real part of $t(z)$ is constant $t_0$, but its imaginary part will increase with $z$. Then the imginary part of $t(z)$ would be constant in $z\in (z_{b,3},+\infty)$, the real part of $t(z)$ will approach to infinity as $z\to +\infty$.

\begin{figure}[htbp]
  \centering
  \includegraphics[width=0.4\textwidth]{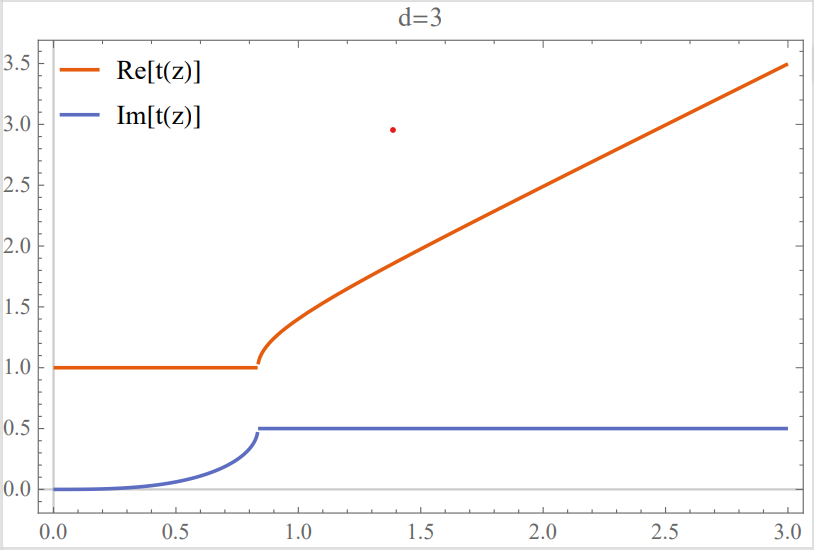}
  \caption{Plot of the RT surface $t(z)$ for real $z$ in the $d=3$ case. }
  \label{path_fig}
\end{figure}

For $d=4$ it is easy to see that the RT surface $t(z)$ is real for $z\in(\delta,+\infty)$. Thus the RT surface in this region will explore the Lorentzian metric and approach to the horizon. 

\section{III: Higher-Dimensional Strip  in the Black Hole}

In this section, we provide details of the calculation for the RT surface in a higher-dimensional black hole. We set $z_h=1$ and assume the strip satisfies $t_0\ll 1$. This allows us to perturbatively compute the area of the RT surface for the timelike strip. Consider a strip $s_E(0,0;\tau_0,0)$ in Euclidean QFTs with $\tau_0\ll 1$. The turning point $z_{\tau_0}$ can be obtained using (\ref{Euclidean_condition}) 
, along with the condition
\bea
\int_{0}^{z_{\tau_0}}\tau'(z)dz =\frac{\tau_0}{2}.
\eea
One could perturbatively solve $z_{\tau_0}$ as
\bea\label{ztau0_bh}
z_{\tau_0}=\zeta_0 \tau_0+ \zeta_d \tau_0^{d+1},
\eea
with 
\bea
&&\zeta_0=\frac{\Gamma(\frac{1}{2(d-1)})}{2\sqrt{\pi}\Gamma(\frac{d}{2(d-1)})},\nn\\
&&\zeta_d=-\frac{\zeta_0^{d+1} \left(2 d-2+\frac{(d-3) \Gamma \left(\frac{1}{2 (d-1)}\right) \Gamma \left(\frac{d}{d-1}\right)}{\Gamma \left(\frac{1-3 d}{2-2 d}\right) \Gamma \left(\frac{d}{2 (d-1)}\right)}\right)}{4 (d-1)^2}.
\eea
With the analytical continuation $\tau_0 \to i t_0$, the turning point $z_{t_0}$ can be determined.The RT surface $t=t(z)$ can also be obtained perturbatively. The conserved constant  $p_t$ can be obtained by using
\bea
f(z_t) z_t^{2(1-d)}+p_t^2=0.
\eea
From (\ref{area_term}) we will have
\bea\label{area_function_bh}
\mathcal{A}=2\int_{C}dz \mathcal{L} \ \text{with}\ \frac{1}{z^{d-1}\sqrt{f(z)+p_t^2z^{2(d-1)}}},
\eea
where $C$ is the path connecting $z=\delta$ to $z_t$. One can choose the path $C$ as in the vacuum case. A more practical approach to evaluating the integral is to first obtain the result for the Euclidean case and then apply analytical continuation.

The area for the RT surface in the Euclidean metric is given by
\bea
\mathcal{A}_E=2\int_{\delta}^{z_{\tau_0}}dz\frac{1}{z^{d-2}} \sqrt{\frac{1}{z^{2 d} z_{\tau_0}^{2-2 d} \left(1-z_{\tau_0}^d\right)-z^{d+2}+z^2}}.\nn
\eea
One could work out this integration perturbatively 
\bea
\mathcal{A}_E=\frac{2}{(d-2)\delta^{d-2}}-\frac{2\sqrt{\pi } \Gamma \left(\frac{d}{2 (d-1)}\right)}{ z_{\tau_0}^{d-2} (d-2) \Gamma \left(\frac{1}{2 (d-1)}\right)}+\frac{1}{2} \sqrt{\pi } z_{\tau_0}^2 \left[\frac{(d-3) \Gamma \left(\frac{d}{d-1}\right)}{(d-1) \Gamma \left(\frac{d+1}{2 (d-1)}\right)}+\frac{4 \Gamma \left(\frac{3 d-2}{2 (d-1)}\right)}{d \Gamma \left(\frac{1}{2 (d-1)}\right)}\right].\nn
\eea
Substituting (\ref{ztau0_bh}) into the above expression, we can obtain the area of the RT surface in the Euclidean metric. By applying the continuation $\tau_0\to i t_0$, the results in Lorentzian geometry $\mathcal{A}_{bh}(0,0;t_0,0)$ can be derived. The calculations are straightforward, though the expression is lengthy. We will present some results for $d=3,4,5,6$ in Table.\ref{table_results}.

\begin{table}
\begin{tabular}{|c|c|c|c|}
  \hline
  d & $\mathcal{A}_{bh}(0,0;t_0,0)$ & $\mathcal{A}_{bh}(0,0;0,x_0)$& $\beta_d/\alpha_d$ \\
  \hline
  3 & $\frac{2}{\delta }+\frac{i \left(8 \pi ^3\right)}{t_0 \Gamma \left(\frac{1}{4}\right)^4}$ & $\frac{2}{\delta }-\frac{8 \pi ^3}{x_0 \Gamma \left(\frac{1}{4}\right)^4}+\frac{x_0^2 \Gamma \left(\frac{1}{4}\right)^4}{64 \pi ^2}$ & 0\\
  \hline
  4 & $\frac{1}{ \delta ^2}+\frac{4 \pi ^{3/2} \Gamma \left(\frac{2}{3}\right)^3}{t_0^2 \Gamma \left(\frac{1}{6}\right)^3}-\frac{t_0^2 \Gamma \left(\frac{1}{6}\right)^2 \Gamma \left(\frac{7}{6}\right)^2}{20 \sqrt{2} \pi ^2 \Gamma \left(\frac{5}{3}\right)}$ & $\frac{1}{\delta ^2}-\frac{\pi ^{3/2} \Gamma \left(\frac{2}{3}\right) \Gamma \left(\frac{5}{3}\right)^2}{4 x_0^2 \left(\Gamma \left(\frac{1}{6}\right) \Gamma \left(\frac{7}{6}\right)^2\right)}+\frac{2 \sqrt{\pi } x_0^2 \Gamma \left(\frac{1}{3}\right) \Gamma \left(\frac{7}{6}\right)}{15 \Gamma \left(\frac{5}{6}\right)^2 \Gamma \left(\frac{5}{3}\right)^2}$ & $-\frac{1}{3}$ \\
  \hline
  5& $\frac{2}{3 \delta ^3}-\frac{i \pi ^2 \Gamma \left(\frac{13}{8}\right) \Gamma \left(\frac{5}{8}\right)^3}{480 t^3 \Gamma \left(\frac{9}{8}\right)^4}-\frac{t^2 \Gamma \left(\frac{1}{8}\right)^2 \Gamma \left(\frac{9}{8}\right)}{12\ 2^{3/4} \pi  \Gamma \left(\frac{5}{8}\right) \Gamma \left(\frac{3}{4}\right)}$ & $\frac{2}{3 \delta ^3}-\frac{\pi ^2 \Gamma \left(\frac{13}{8}\right) \Gamma \left(\frac{5}{8}\right)^3}{480 x_0^3 \Gamma \left(\frac{9}{8}\right)^4}+\frac{10 \sqrt[4]{2} x_0^2 \Gamma \left(\frac{9}{8}\right)^3}{3 \pi  \Gamma \left(\frac{3}{4}\right) \Gamma \left(\frac{13}{8}\right)}$ &$-\frac{1}{2}$\\
\hline
6& $\frac{1}{2 \delta ^4}-\frac{\pi ^{5/2} \Gamma \left(\frac{8}{5}\right)^5}{972 t_0^4 \Gamma \left(\frac{11}{10}\right)^5}-\frac{27 t_0^2 \Gamma \left(\frac{1}{10}\right) \Gamma \left(\frac{11}{10}\right) \Gamma \left(\frac{6}{5}\right)}{200 \sqrt{\pi } \Gamma \left(\frac{8}{5}\right)^2 \Gamma \left(\frac{17}{10}\right)}$ & $\frac{1}{2 \delta ^4}-\frac{\pi ^{5/2} \Gamma \left(\frac{8}{5}\right)^5}{972 x_0^4 \Gamma \left(\frac{11}{10}\right)^5}+\frac{45 x_0^2 \Gamma \left(\frac{6}{5}\right) \Gamma \left(\frac{11}{10}\right)^2}{14 \sqrt{\pi } \Gamma \left(\frac{7}{10}\right) \Gamma \left(\frac{8}{5}\right)^2}$& $-\frac{3}{5}$\\
\hline 
\end{tabular}
\caption{List of the perturbative results for $\mathcal{A}_{bh}(0,0;t_0,0)$ and $\mathcal{A}_{bh}(0,0;0,x_0)$ for $d=3,4,5,6$.}
\label{table_results}
\end{table}
Similarly, one could calculate the RT surface for spacelike strip $s(0,0;0,x_0)$. Assume $x_0\ll 1$ the results can be obtained perturbatively. The turning point $z_x$ could be solved by using (\ref{Euclidean_condition}) with $p_\tau=0$ and the condition $\int_{0}^{z_x}dz x'(z)=\frac{x_0}{2}$. The area can be calculated by
\bea
\mathcal{A}_s=2\int_{\delta}^{z_x}dz \frac{1}{z^{d-2}} \sqrt{\frac{1}{\left(z^d-1\right) \left(z^{2 d} z_x^{2-2 d}-z^2\right)}}.
\eea
We show the results for $d=3,4,5,6$ in Table.\ref{table_results}.

\section{IV: Derivation of Eq.~(12) in the main text}
Here we present a detailed derivation of Eq.~(12) in the main text for the BTZ black hole case.  
In the main text, the inside and outside contributions to the timelike RT surface for the interval $[0,t_0]$ are
\begin{align}
 \mathcal{A}_{\mathrm{in}} &= 2\log\!\left[(1+\coth\frac{t_0}{2z_h})\sinh\frac{t_0}{2z_h} \right] + i\pi, \nonumber\\
 \mathcal{A}_{\mathrm{out}} &= 2\log\!\left[\frac{2z_h}{\delta}\,\frac{1}{1+\coth\frac{t_0}{2z_h}} \right],
\end{align}
where $\delta$ is the UV cutoff.

Thus, the total area of the timelike RT surface is
\begin{align}
\mathcal{A}(0,0;t_0,0) = \mathcal{A}_{\mathrm{in}} + \mathcal{A}_{\mathrm{out}}
= 2\log\!\left[\frac{2z_h}{\delta}\sinh\frac{t_0}{2z_h} \right] + i\pi.
\end{align}

In general, for any interval $(t, x; t', x')$, we have
\begin{align}
\mathcal{A}(t, x; t', x') 
&= \log\!\left[\frac{2z_h}{\delta}\sinh\frac{-(u-u'-i\epsilon)}{2z_h} \right] 
+ \log\!\left[\frac{2z_h}{\delta}\sinh\frac{v-v'-i\epsilon}{2z_h} \right],
\end{align}
where $u = t-x$, $v = t+x$, $u' = t'-x'$ and $v' = t'+x'$.  
The above expression is valid for both spacelike and timelike intervals.

For the spacelike intervals $(0, -t_0; 0, 0)$ and $(0, 0; 0, t_0)$, we obtain
\begin{align}
\mathcal{A}(0, -t_0; 0, 0) &= \mathcal{A}(0, 0; 0, t_0) \nonumber\\
&= \log\!\left[\frac{2z_h}{\delta}\sinh\frac{t_0}{2z_h} \right] - i\pi 
 + \log\!\left[\frac{2z_h}{\delta}\sinh\frac{t_0}{2z_h} \right] + i\pi \nonumber\\
&= 2\log\!\left[\frac{2z_h}{\delta}\sinh\frac{t_0}{2z_h} \right].
\end{align}

The partial derivative of $\mathcal{A}$ with respect to $t$ reads
\begin{align}
\partial_t \mathcal{A}(t, x; t', x') 
= -\frac{1}{2 z_h} \coth\!\left( \frac{-(u - u' - i\epsilon)}{2 z_h} \right)
+ \frac{1}{2 z_h} \coth\!\left( \frac{v - v' - i\epsilon}{2 z_h} \right).
\end{align}
For the spacelike intervals $(0, x; 0, 0)$ this becomes
\begin{align}
\partial_t \mathcal{A}(0, x; 0, 0)
&= -\frac{1}{2 z_h} \frac{\cosh\!\left( \frac{x+ i\epsilon}{2 z_h} \right)}{\sinh\!\left( \frac{x+ i\epsilon}{2 z_h} \right)}
+ \frac{1}{2 z_h} \frac{\cosh\!\left( \frac{x- i\epsilon}{2 z_h} \right)}{\sinh\!\left( \frac{x- i\epsilon}{2 z_h} \right)} \nonumber\\
&= \frac{1}{2 z_h} \frac{\cosh\!\left( \frac{x- i\epsilon}{2 z_h} \right)\sinh\!\left( \frac{x+ i\epsilon}{2 z_h} \right) 
- \cosh\!\left( \frac{x+ i\epsilon}{2 z_h} \right)\sinh\!\left( \frac{x- i\epsilon}{2 z_h} \right)}
{\sinh\!\left( \frac{x- i\epsilon}{2 z_h} \right) \sinh\!\left( \frac{x+ i\epsilon}{2 z_h} \right)} \nonumber\\
&= \frac{1}{2 z_h} \frac{\sinh\!\left( \frac{i\epsilon}{ z_h} \right)}
{\sinh\!\left( \frac{x- i\epsilon}{2 z_h} \right) \sinh\!\left( \frac{x+ i\epsilon}{2 z_h} \right)}.
\end{align}

For $x \neq 0$, the denominator is finite, while the numerator vanishes as $\epsilon \to 0$, hence $\partial_t\mathcal{A} = 0$.  
For $x \to 0$, we have
\begin{align}
\p_t\mathcal{A}(0, x; 0, 0)
&=\frac{1}{2 z_h}\frac{\sinh\left( \frac{i\epsilon}{ z_h} \right)}{\sinh\left( \frac{x- i\epsilon}{2 z_h} \right) \sinh\left( \frac{x+ i\epsilon}{2 z_h} \right)}\nn\\
&=\frac{1}{2 z_h}\frac{\frac{i\epsilon}{ z_h}}{\frac{x- i\epsilon}{2 z_h}\frac{x+ i\epsilon}{2 z_h}}\nn\\
&=\frac{1}{x- i\epsilon}-\frac{1}{x+ i\epsilon},
\end{align}
Using the Sokhotski–Plemelj formula
\begin{align}
\frac{1}{x \mp i\epsilon} = \mathcal{P}(\frac{1}{x})\pm i\pi\delta(x),
\end{align}
we obtain
\begin{align}
\partial_t \mathcal{A}(0, x; 0, 0) = 2\pi i\,\delta(x).
\end{align}
Therefore,
\begin{align}
\int_{-t_0}^{t_0} dx\, \partial_t \mathcal{A}(0, x; 0, 0) = 2\pi i.
\end{align}

Finally,
\begin{align}
&\frac{1}{2}\left[ \mathcal{A}(0,-t_0;0,0) + \mathcal{A}(0,0;0,t_0) \right]
+ \frac{1}{2} \int_{-t_0}^{t_0} dx\, \partial_t \mathcal{A}(0,x;0,0) \nonumber\\
&= 2\log\!\left[\frac{2z_h}{\delta}\sinh\frac{t_0}{2z_h} \right] + i\pi 
= \mathcal{A}(0,0;t_0,0) = \mathcal{A}_{\mathrm{in}} + \mathcal{A}_{\mathrm{out}},
\end{align}
which directly yields
\begin{align}
\mathcal{A}_{\mathrm{in}} 
= \frac{1}{2}\left[ \mathcal{A}(0,-t_0;0,0) + \mathcal{A}(0,0;0,t_0) \right]
+ \frac{1}{2} \int_{-t_0}^{t_0} dx\, \partial_t \mathcal{A}(0,x;0,0)
- \mathcal{A}_{\mathrm{out}}.
\end{align}
This completes the proof of Eq.~(12) in the main text.

\section{V: Derivation of Eq.~(21) in the main text}

Here we present a detailed derivation of Eq.~(21) in the main text. In the vacuum case, we have
\begin{align}\label{hd_v}
\mathcal{A}(0,0;t_0,0)
=\frac{1}{2}\left[\mathcal{A}(0,-t_0;0,0)+\mathcal{A}(0,0;0,t_0)\right]
+I,
\end{align}
where
\begin{align}
I\equiv i\frac{(-i)^{\,d-2}-1}{(d-2)\pi}\,t_0^{\,d-2}
\int_{-t_0}^{t_0}dx\,x^{d-2}\,\partial_t \mathcal{A}(0,x;0,0).
\end{align}
For the black hole background, the spacelike and timelike surfaces receive thermal corrections given by
\begin{align}
\mathcal{A}_{bh}(0,0;0,x)&=\mathcal{A}(0,0;0,x)+\delta \mathcal{A}_{bh}(0,0;0,x),\qquad \delta \mathcal{A}_{bh}(0,0;0,x)=\alpha_d x^2+O(x^{2+d}),\\
\mathcal{A}_{bh}(0,0;t_0,0)&=\mathcal{A}(0,0;t_0,0)+\delta \mathcal{A}_{bh}(0,0;t_0,0) ,\qquad \delta \mathcal{A}_{bh}(0,0;t_0,0)=\beta_d\,t_0^2+O(t_0^{2+d}),
\end{align}
with the universal ratio
\begin{align}
\frac{\beta_d}{\alpha_d}=-\,\frac{d-3}{d-1}.
\end{align}
Since the first time derivative $\partial_t \mathcal{A}(0,x;0,0)$ in $t\to0$ is not affected by thermal corrections to the order considered, we can replace $\mathcal{A}$ by $\mathcal{A}_{bh}-\delta \mathcal{A}_{bh}$ in the vacuum relation without modifying $I$. Substituting into (\ref{hd_v}) gives
\begin{align}
\mathcal{A}_{bh}(0,0;t_0,0)-\delta \mathcal{A}_{bh}(0,0;t_0,0)
&=\frac12\left[\mathcal{A}_{bh}(0,-t_0;0,0)-\delta \mathcal{A}_{bh}(0,-t_0;0,0)
+\mathcal{A}_{bh}(0,0;0,t_0)-\delta \mathcal{A}_{bh}(0,0;0,t_0)\right]+I.
\end{align}
Using $\delta \mathcal{A}_{bh}(0,-t_0;0,0)=\delta \mathcal{A}_{bh}(0,0;0,t_0)=\alpha_d t_0^2$ and $\delta \mathcal{A}_{bh}(0,0;t_0,0)=\beta_d t_0^2$, we obtain
\begin{align}
\mathcal{A}_{bh}(0,0;t_0,0)
&=\frac12\left[\mathcal{A}_{bh}(0,-t_0;0,0)+\mathcal{A}_{bh}(0,0;0,t_0)\right]
-\frac12\left[\delta \mathcal{A}_{bh}(0,-t_0;0,0)+\delta \mathcal{A}_{bh}(0,0;0,t_0)\right]
+\beta_d\,t_0^2+I\nn\\
&=\frac12\left[\mathcal{A}_{bh}(0,-t_0;0,0)+\mathcal{A}_{bh}(0,0;0,t_0)\right]
+\left(\frac{\beta_d}{2\alpha_d}-\frac12\right)\left[\delta \mathcal{A}_{bh}(0,-t_0;0,0)+\delta \mathcal{A}_{bh}(0,0;0,t_0)\right]
+I.
\end{align}
Finally, using $\beta_d/\alpha_d=-(d-3)/(d-1)$, the coefficient becomes
\begin{align}
\frac{\beta_d}{2\alpha_d}-\frac12
=-\frac{d-3}{2(d-1)}-\frac{d-1}{2(d-1)}
=-\frac{d-2}{d-1},
\end{align}
leading to
\begin{align}
\mathcal{A}_{bh}(0,0;t_0,0)
=\frac12\left[\mathcal{A}_{bh}(0,-t_0;0,0)+\mathcal{A}_{bh}(0,0;0,t_0)\right]
-\frac{d-2}{d-1}\left[\delta \mathcal{A}_{bh}(0,-t_0;0,0)+\delta \mathcal{A}_{bh}(0,0;0,t_0)\right]
+I,
\end{align}
which is exactly Eq.~(21) in the main text.


\begin{thebibliography}{00}

\bibitem{Stephens:1993an}
C.~R.~Stephens, G.~'t Hooft and B.~F.~Whiting,
``Black hole evaporation without information loss,''
Class. Quant. Grav. \textbf{11}, 621-648 (1994)
[arXiv:gr-qc/9310006 [gr-qc]].
\bibitem{Susskind:1993if}
L.~Susskind, L.~Thorlacius and J.~Uglum,
``The Stretched horizon and black hole complementarity,''
Phys. Rev. D \textbf{48}, 3743-3761 (1993)
[arXiv:hep-th/9306069 [hep-th]].

\bibitem{Ryu:2006bv}
S.~Ryu and T.~Takayanagi,
``Holographic derivation of entanglement entropy from AdS/CFT,''
Phys. Rev. Lett. \textbf{96}, 181602 (2006)
[arXiv:hep-th/0603001 [hep-th]].
\bibitem{Hubeny:2007xt}
V.~E.~Hubeny, M.~Rangamani and T.~Takayanagi,
``A Covariant holographic entanglement entropy proposal,''
JHEP \textbf{07}, 062 (2007)
[arXiv:0705.0016 [hep-th]].

\bibitem{Hubeny:2012ry}
V.~E.~Hubeny,
``Extremal surfaces as bulk probes in AdS/CFT,''
JHEP \textbf{07} (2012), 093
[arXiv:1203.1044 [hep-th]].



\bibitem{Engelhardt:2013tra}
N.~Engelhardt and A.~C.~Wall,
``Extremal Surface Barriers,''
JHEP \textbf{03}, 068 (2014)
[arXiv:1312.3699 [hep-th]].

\bibitem{Hartman:2013qma}
T.~Hartman and J.~Maldacena,
``Time Evolution of Entanglement Entropy from Black Hole Interiors,''
JHEP \textbf{05} (2013), 014
[arXiv:1303.1080 [hep-th]].

\bibitem{Fidkowski:2003nf}
L.~Fidkowski, V.~Hubeny, M.~Kleban and S.~Shenker,
``The Black hole singularity in AdS / CFT,''
JHEP \textbf{02} (2004), 014
[arXiv:hep-th/0306170 [hep-th]].

\bibitem{Frenkel:2020ysx}
A.~Frenkel, S.~A.~Hartnoll, J.~Kruthoff and Z.~D.~Shi,
``Holographic flows from CFT to the Kasner universe,''
JHEP \textbf{08} (2020), 003
[arXiv:2004.01192 [hep-th]].

\bibitem{Stanford:2014jda}
D.~Stanford and L.~Susskind,
``Complexity and Shock Wave Geometries,''
Phys. Rev. D \textbf{90}, no.12, 126007 (2014)
[arXiv:1406.2678 [hep-th]].

\bibitem{Brown:2015bva}
A.~R.~Brown, D.~A.~Roberts, L.~Susskind, B.~Swingle and Y.~Zhao,
``Holographic Complexity Equals Bulk Action?,''
Phys. Rev. Lett. \textbf{116}, no.19, 191301 (2016)
[arXiv:1509.07876 [hep-th]].

\bibitem{Doi:2022iyj}
K.~Doi, J.~Harper, A.~Mollabashi, T.~Takayanagi and Y.~Taki,
``Pseudoentropy in dS/CFT and Timelike Entanglement Entropy,''
Phys. Rev. Lett. \textbf{130}, no.3, 031601 (2023)
[arXiv:2210.09457 [hep-th]].

\bibitem{Nakata:2020luh}
Y.~Nakata, T.~Takayanagi, Y.~Taki, K.~Tamaoka and Z.~Wei,
``New holographic generalization of entanglement entropy,''
Phys. Rev. D \textbf{103} (2021) no.2, 026005
doi:10.1103/PhysRevD.103.026005
[arXiv:2005.13801 [hep-th]].

\bibitem{Olson:2010jy}
S.~J.~Olson and T.~C.~Ralph,
``Entanglement between the future and past in the quantum vacuum,''
Phys. Rev. Lett. \textbf{106} (2011), 110404
[arXiv:1003.0720 [quant-ph]].

\bibitem{Cotler:2017anu}
J.~Cotler, C.~M.~Jian, X.~L.~Qi and F.~Wilczek,
``Superdensity Operators for Spacetime Quantum Mechanics,''
JHEP \textbf{09} (2018), 093
[arXiv:1711.03119 [quant-ph]].

\bibitem{Wang:2018jva}
P.~Wang, H.~Wu and H.~Yang,
``Fix the dual geometries of $T\bar{T}$ deformed CFT$_2$ and highly excited states of CFT$_2$,''
Eur. Phys. J. C \textbf{80} (2020) no.12, 1117
[arXiv:1811.07758 [hep-th]].
\bibitem{Lerose:2021svg}
A.~Lerose, M.~Sonner and D.~A.~Abanin,
``Scaling of temporal entanglement in proximity to integrability,''
Phys. Rev. B \textbf{104} (2021) no.3, 035137
[arXiv:2104.07607 [quant-ph]].

\bibitem{Giudice:2021smd}
G.~Giudice, G.~Giudici, M.~Sonner, J.~Thoenniss, A.~Lerose, D.~A.~Abanin and L.~Piroli,
``Temporal Entanglement, Quasiparticles, and the Role of Interactions,''
Phys. Rev. Lett. \textbf{128} (2022) no.22, 220401
[arXiv:2112.14264 [cond-mat.stat-mech]].


\bibitem{Narayan:2022afv}
K.~Narayan,
``de Sitter space, extremal surfaces, and time entanglement,''
Phys. Rev. D \textbf{107} (2023) no.12, 126004
doi:10.1103/PhysRevD.107.126004
[arXiv:2210.12963 [hep-th]].

\bibitem{Liu:2022ugc}
B.~Liu, H.~Chen and B.~Lian,
``Entanglement entropy of free fermions in timelike slices,''
Phys. Rev. B \textbf{110} (2024) no.14, 144306
[arXiv:2210.03134 [cond-mat.stat-mech]].

\bibitem{Glorioso:2024xan}
P.~Glorioso, X.~L.~Qi and Z.~Yang,
``Space-time generalization of mutual information,''
JHEP \textbf{05} (2024), 338
[arXiv:2401.02475 [quant-ph]].

\bibitem{Milekhin:2025ycm}
A.~Milekhin, Z.~Adamska and J.~Preskill,
``Observable and computable entanglement in time,''
[arXiv:2502.12240 [quant-ph]].

\bibitem{Li:2022tsv}
Z.~Li, Z.~Q.~Xiao and R.~Q.~Yang,
JHEP \textbf{04} (2023), 004
doi:10.1007/JHEP04(2023)004
[arXiv:2211.14883 [hep-th]].
\bibitem{Doi:2023zaf}
K.~Doi, J.~Harper, A.~Mollabashi, T.~Takayanagi and Y.~Taki,
``Timelike entanglement entropy,''
JHEP \textbf{05}, 052 (2023)
[arXiv:2302.11695 [hep-th]].

\bibitem{Heller:2024whi}
M.~P.~Heller, F.~Ori and A.~Serantes,
``Geometric Interpretation of Timelike Entanglement Entropy,''
Phys. Rev. Lett. \textbf{134} (2025) no.13, 131601
[arXiv:2408.15752 [hep-th]].

\bibitem{Anegawa:2024kdj}
T.~Anegawa and K.~Tamaoka,
``Black hole singularity and timelike entanglement,''
JHEP \textbf{10} (2024), 182
[arXiv:2406.10968 [hep-th]].

\bibitem{Guo:2024lrr}
W.~z.~Guo, S.~He and Y.~X.~Zhang,
``Relation between timelike and spacelike entanglement entropy,''
[arXiv:2402.00268 [hep-th]].


\bibitem{Heemskerk:2012mn}
I.~Heemskerk, D.~Marolf, J.~Polchinski and J.~Sully,
``Bulk and Transhorizon Measurements in AdS/CFT,''
JHEP \textbf{10} (2012), 165
[arXiv:1201.3664 [hep-th]].

\bibitem{Susskind:2012uw}
L.~Susskind,
``The Transfer of Entanglement: The Case for Firewalls,''
[arXiv:1210.2098 [hep-th]].

\bibitem{Guo:2025mwp}
W.~z.~Guo,
``Measuring the Black Hole Interior from the Exterior,''
[arXiv:2505.09878 [hep-th]].
\bibitem{Czech:2012bh}
  B.~Czech, J.~L.~Karczmarek, F.~Nogueira and M.~Van Raamsdonk,
  ``The Gravity Dual of a Density Matrix,''
  Class.\ Quant.\ Grav.\  {\bf 29}, 155009 (2012)
  [arXiv:1204.1330 [hep-th]].
\bibitem{Jafferis:2015del}
  D.~L.~Jafferis, A.~Lewkowycz, J.~Maldacena and S.~J.~Suh,
``Relative entropy equals bulk relative entropy,''
JHEP \textbf{06} (2016), 004
[arXiv:1512.06431 [hep-th]].

\bibitem{Dong:2016eik}
X.~Dong, D.~Harlow and A.~C.~Wall,
``Reconstruction of Bulk Operators within the Entanglement Wedge in Gauge-Gravity Duality,''
Phys. Rev. Lett. \textbf{117} (2016) no.2, 021601
[arXiv:1601.05416 [hep-th]].


\bibitem{Xu:2024yvf}
J.~Xu and W.~z.~Guo,
``Imaginary part of timelike entanglement entropy,''
JHEP \textbf{02} (2025), 094
[arXiv:2410.22684 [hep-th]].

\bibitem{Kasner:1921zz}
E.~Kasner,
``Geometrical theorems on Einstein's cosmological equations,''
Am. J. Math. \textbf{43} (1921), 217-221

\bibitem{Belinski:1973zz}
V.~A.~Belinski and I.~M.~Khalatnikov,
``Effect of Scalar and Vector Fields on the Nature of the Cosmological Singularity,''
Sov. Phys. JETP \textbf{36} (1973), 591

\bibitem{Penington:2019npb}
G.~Penington,
``Entanglement Wedge Reconstruction and the Information Paradox,''
JHEP \textbf{09} (2020), 002
[arXiv:1905.08255 [hep-th]].
\bibitem{Almheiri:2019psf}
A.~Almheiri, N.~Engelhardt, D.~Marolf and H.~Maxfield,
``The entropy of bulk quantum fields and the entanglement wedge of an evaporating black hole,''
JHEP \textbf{12}, 063 (2019)
[arXiv:1905.08762 [hep-th]].

\bibitem{Almheiri:2019hni}
A.~Almheiri, R.~Mahajan, J.~Maldacena and Y.~Zhao,
``The Page curve of Hawking radiation from semiclassical geometry,''
JHEP \textbf{03}, 149 (2020)
[arXiv:1908.10996 [hep-th]].

\bibitem{Almheiri:2020cfm}
A.~Almheiri, T.~Hartman, J.~Maldacena, E.~Shaghoulian and A.~Tajdini,
``The entropy of Hawking radiation,''
Rev. Mod. Phys. \textbf{93}, no.3, 035002 (2021)
[arXiv:2006.06872 [hep-th]].




\end{thebibliography}

\end{document}